\documentclass[sigplan,10pt]{acmart}
\renewcommand\footnotetextcopyrightpermission[1]{}
\usepackage{algorithm}
\usepackage{algpseudocode}
\usepackage{etoolbox}
\newcommand{\savebodyfloatcommand}[1]{\csletcs{bodyfloat#1}{#1}}
\forcsvlist{\savebodyfloatcommand}{@dbflt,@dblfloatplacement,@addtodblcol,@addtocurcol,@addtotoporbot,@addtobot,@doclearpage,@combinedblfloats}
\usepackage{stfloats}
\newcommand{\saveappendixfloatcommand}[1]{%
  \csletcs{appendixfloat#1}{#1}\csletcs{#1}{bodyfloat#1}}
\forcsvlist{\saveappendixfloatcommand}{@dbflt,@dblfloatplacement,@addtodblcol,@addtocurcol,@addtotoporbot,@addtobot,@doclearpage,@combinedblfloats}
\newcommand{\enableappendixfloatcommand}[1]{\csletcs{#1}{appendixfloat#1}}
\usepackage{graphicx}
\usepackage{subfig}
\usepackage{subcaption}

\usepackage{xspace}
\usepackage{enumitem}
\usepackage{url}

\setcopyright{none}

\newcommand{\sys}{\textsf{Duet}\xspace}
\newcommand{\evalfigdir}{figures-duet}

\newtheorem{invariant}{Invariant}

\newtheorem{observation}{Observation}

\algrenewcommand\algorithmicindent{2.0em}

\newcommand{\myindent}{\hspace{\algorithmicindent}}

\newcommand{\msgtag}[1]{\textsc{{#1}}}

\newcommand{\msg}[2]{\ensuremath{\langle\msgtag{#1},#2\rangle}}

\newcommand{\best}{\textsc{tree}\xspace}
\newcommand{\propose}{\textsc{propose}\xspace}
\newcommand{\proposal}{\textsc{proposal}\xspace}
\newcommand{\prevote}{\textsc{prevote}\xspace}
\newcommand{\precommit}{\textsc{precommit}\xspace}

\newcommand{\tmr}{\ensuremath{tmr}}
\newcommand{\peer}{\texttt{peer}}
\newcommand{\children}{\texttt{children}}

\newcommand{\height}{\ensuremath{h}}

\definecolor{mylightgray}{gray}{0.8}
\definecolor{mydarkgray}{gray}{0.6}
\definecolor{offline}{gray}{0.7}
\newcommand{\algemph}[1]{\colorbox{mylightgray}{#1}}

\newcommand{\argmin}{\arg\!\min}

\begin{document}

\title{\sys: Co-Optimizing P2P Message Propagation and Rotating-Leader Consensus}

\author{Yifeng Ye}
\affiliation{%
  \institution{Shanghai Jiao Tong University}
  \country{China}
}
\author{Rongji Huang}
\affiliation{%
  \institution{Shanghai Academy of Future Internet Technology}
  \country{China}
}
\author{Gerui Wang}
\affiliation{%
  \institution{Beijing Academy of Blockchain and Edge Computing}
  \country{China}
}
\author{Mingchao Wan}
\affiliation{%
  \institution{Beijing Academy of Blockchain and Edge Computing}
  \country{China}
}
\author{Yuxing Duan}
\affiliation{%
  \institution{Beijing Academy of Blockchain and Edge Computing}
  \country{China}
}
\author{Jingjing Zhang}
\affiliation{%
  \institution{Fudan University}
  \country{China}
}
\author{Shengyun Liu}
\affiliation{%
  \institution{Shanghai Jiao Tong University}
  \country{China}
}

\begin{abstract}

In blockchain systems, peer-to-peer (P2P) overlay networks play a crucial role in providing reliable, scalable and efficient message-delivery services to upper layers.
However, the consensus layer and the underlying P2P network remain mutually opaque in existing blockchains, waiving the opportunity for further improvement.
Unlike many other P2P applications, a blockchain can naturally be modeled as a state machine.
We leverage this abstraction to maintain network topology and latency information in a trusted and coordinated manner, making such information available for cross-layer optimization.
Building on this capability, we propose three improvements to rotating-leader consensus protocols and their underlying P2P networks: (1) accelerating leader rotation by optimizing the proposer sequence; (2) introducing a new reliable-broadcast paradigm that uses efficient tree-based dissemination in the common case while falling back to per-node gossip when necessary; and (3) constructing latency-aware dissemination trees to reduce propagation latency and communication overhead.
We integrate these techniques into Tendermint and libp2p and evaluate our prototype on Amazon EC2 using up to 300 nodes distributed across 10 geographically dispersed regions. 
Our results show that, compared with gossip-based dissemination over the same network topology, our prototype improves peak throughput by up to \(7.26\times\).

\end{abstract}

\maketitle
\pagestyle{plain}

\section{Introduction}

Blockchain systems, particularly permissionless ones that enable open participation by widely-scattered players, are typically built on peer-to-peer (P2P) overlay networks.
P2P networks provide upper layers with reliable, scalable, and efficient message dissemination, ensuring that every message eventually reaches all correct nodes.
These properties are essential for system scalability and simplify the design, implementation, and maintenance of upper-layer consensus protocols, which establish agreement among nodes on the ordering of blocks.

Compared with all-to-all communication, P2P networks allow each node to maintain connections with and forward messages to only a small subset of other nodes, substantially reducing communication complexity while providing reliable message delivery with high probability. 
Hence, P2P networking combined with gossip-based message dissemination has become the de facto approach to scaling blockchain networks. 
For example, Bitcoin~\cite{bitcoin} uses an unstructured P2P network to connect nodes, whereas Ethereum~\cite{ethereum} employs a Kademlia-based~\cite{Kademlia} structured overlay for node discovery. 
Other permissionless blockchains, including XRP~\cite{xrp}, Filecoin~\cite{filecoin}, which builds on IPFS~\cite{ipfs}, Algorand~\cite{algorand,algorandp2p}, and Avalanche~\cite{rocket2019scalable}, also rely on P2P networks. 
Permissioned blockchains, such as Hyperledger Fabric~\cite{fabric}, similarly leverage P2P networks to support large-scale deployments~\cite{fabricp2p}.

By providing a reliable broadcast primitive~\cite{cachin2011introduction}, P2P network largely simplifies the design and implementation of the upper consensus layer. 
For instance, Nakamoto consensus~\cite{bitcoin} follows a very simple rule in which honest nodes extend the longest chain they are aware of, which represents the greatest cumulative proof-of-work effort.
Tendermint~\cite{tendermint} relies solely on a three-phase message exchange to both reach agreement and rotate proposers, even in the presence of faulty nodes.
HashGraph~\cite{hashgraph} reaches consensus through a gossip-about-gossip protocol, in which nodes merely disseminate transactions together with their communication histories, allowing each node to determine consensus locally through virtual voting.
In contrast, traditional Byzantine fault-tolerant (BFT) consensus protocols~\cite{pbft,zyzzyva,sbft,hotstuff} are designed over point-to-point communication, requiring explicit view-change and view-synchronization mechanisms~\cite{cogsworth,fever,squad} to replace faulty proposers and keep nodes synchronized, as well as substantial protocol complexity to reduce communication overhead~\cite{hbbft}.

Napster, launched in 1999, marked the first generation of P2P networks~\cite{napster}.
Since then, P2P overlays have been extensively studied in both academia and industry~\cite{chord,ODRI,Pastry,CAN,Viceroy,kelips}.
Prior to blockchain, P2P networks were primarily used for decentralized content sharing and communication~\cite{Gnutella,bittorrent}, generally without a centralized coordination mechanism or equivalent functionality.

Blockchain systems present a fundamentally different setting.
They can naturally be modeled as replicated state machines~\cite{smr}, providing logically centralized, yet physically decentralized, consensus and contract-execution layers that can assist in organizing and managing the underlying P2P network.
Meanwhile, in existing blockchain systems, the consensus layer and the underlying P2P network largely remain opaque to each other, leaving opportunities for cross-layer optimization untapped.


In this work, we specifically target rotating-leader consensus (RLC)~\cite{tendermint,hotstuff,syncrlc,banyan,moonshot}, a paradigm that has become prevalent in modern permissioned and permissionless blockchains~\cite{bitcoin,ethereum,combcasper,cosmos}.
In RLC protocols, the leader or proposer rotates across block heights to distribute block-proposal opportunities evenly among nodes.
Rotating-leader consensus avoids relying on a single leader while preserving the simplicity and efficiency of stable-leader consensus.
Moreover, this paradigm allows nodes to reach consensus also on execution results rather than merely on the transactions themselves, thereby effectively mitigating the non-determinism problem~\cite{nondeterminism}.

Although P2P networks offer appealing features, their random topologies and routing paths, together with gossip-based dissemination, still limit the efficiency of high performance blockchain systems.
Tree-based dissemination~\cite{kauri,byzcoin} offers a natural way to reduce the redundancy of gossip-based communication, as each node receives a message along a single dissemination path. 
Applying it to P2P-based rotating-leader consensus, however, raises two challenges.

First, a faulty internal node can isolate all of its descendants by disrupting message forwarding. 
Existing approaches typically address this problem by reconfiguring the tree or falling back to one-to-all communication, both of which assume a fully connected network. 
However, large-scale blockchains operate over P2P overlays, where each node maintains connections to only a small subset of peers. 
Neither approach can therefore be directly applied. 
This calls for fundamentally different recovery mechanisms that account for the connectivity constraints of P2P networks.

Second, fast dissemination to all nodes does not necessarily translate into efficient pipelining.
In a rotating-leader protocol, the proposer at each height can propose a new block only after receiving the preceding blocks (i.e., the prefix).
With a stable leader, this requirement comes for free: the same node proposes every block and already holds the entire prefix. 
Under rotating-leader consensus, however, the proposer changes at every height and must wait for the prefix to arrive before proposing.
Prefix delivery therefore determines how quickly new heights can enter the pipeline, coupling the dissemination tree to the proposer sequence.
A balanced tree may disseminate each block efficiently to all nodes, yet place upcoming proposers several hops from the root, delaying prefix delivery and leaving the pipeline underutilized.

We observe that making network topology and latency information available to the consensus layer opens up new opportunities for jointly optimizing both layers:
\begin{itemize}
    \item Consensus can exploit network information to optimize the proposer sequence, accelerating prefix delivery and improving the throughput of RLC protocols;
    
    \item Consensus votes can serve as implicit acknowledgments of proposal delivery, enabling efficient tree-based dissemination in the common case while retaining gossip-based dissemination as a seamless fallback, thereby establishing a novel reliable broadcast paradigm for P2P networks that we call \textbf{two-stage reliable broadcast};
    
    \item The P2P layer can leverage topology and latency information to construct per-proposer dissemination trees that optimize not only the delivery of each block to all nodes but also prefix delivery to upcoming proposers, while reducing cross-region traffic.
\end{itemize}



We instantiate the above ideas by integrating them into a pipelined variant of Tendermint~\cite{tendermint} running over libp2p~\cite{libp2p}. 
Tendermint is a classical RLC protocol widely adopted by blockchain systems, such as the Cosmos ecosystem~\cite{cosmos}, Binance Chain~\cite{bnb-tendermint}, Bitcoin IPC~\cite{bitcoinipc} and ChainMaker~\cite{flextender}. 
libp2p is a modular P2P networking framework that has seen broad deployment across decentralized systems, including Ethereum's consensus layer~\cite{ethlibp2p} and IPFS~\cite{ipfslibp2p}. 
We refer to our prototype as \sys.
Integrating tree-based dissemination into a P2P-based consensus protocol poses non-trivial challenges, particularly because it invalidates the standard assumption that every correct node receives every message within $\Delta$ after the global stabilization time (GST).
The protocol must therefore be adapted accordingly to preserve liveness and other critical properties.
Another key advantage of our approaches is that, even with inaccurate or stale topology information, \sys maintains liveness through gossip-based dissemination.

We conduct an empirical evaluation on Amazon EC2 using up to 300 nodes distributed across 10 geographically dispersed regions worldwide.
We also compare \sys against gossip-based dissemination and a $K$-ary tree~\cite{kauri}.
The results demonstrate that \sys achieves up to $7.26\times$ higher throughput than gossip-based dissemination over the same topology.

\sys can be readily integrated into permissioned blockchains, which usually provide a configuration module for membership management.
\sys can also be applied to permissionless blockchains by maintaining a network backbone in which nodes may, for example, stake through a smart contract to participate~\cite{lowdegree}.
Although we target RLC protocols, extending the ideas of \sys to multi-leader or DAG-based protocols~\cite{tusk,bullshark,mir,mytumbler,themix,chitu,vegeta} may be even more promising, as they support multiple concurrent proposals and often incur higher bandwidth consumption.

\section{Background}
\subsection{System model}

We consider a system with $N$ nodes $\Pi$ at some time.
We assume all point-to-point links are authenticated and reliable: all messages exchanged between any two correct nodes will eventually arrive.
To prevent message tampering, node $p_i$ utilizes its private key to generate a signature $\sigma_i$ attached to message $m$ it sends.
Upon receiving $m$, the receiver needs to verify $\sigma_i$ by $p_i$'s public key.
Regarding ``deliver'', we mean a node successfully receives and verifies a message $m$.
Since nodes in a P2P network are not fully connected, we adopt a gossip-based protocol for message propagation (i.e., broadcast): For each message $m$, every node except the original sender will forward $m$ to some (specific) neighbors once the node delivers it.
We focus on the following two ideal properties~\cite{cachin2011introduction}:
\begin{itemize}
\item Validity: If a correct node broadcasts a message $m$, then every correct node eventually delivers $m$.
\item Totality: If a correct node delivers a message $m$, then every correct node eventually delivers $m$.
\end{itemize}

Note that P2P networks ensure the above properties with high probability.
We assume a partially synchronous network model~\cite{partialsynchrony}, meaning that after a global stabilization time~(GST), messages exchanged between correct nodes are guaranteed to arrive within a bounded delay $\Delta$.
Inspired by Tendermint's industrial deployment~\cite{cometbft}, we can distinguish two upper bounds, $\Delta_L$ and $\Delta_S$, for large messages (e.g., proposals) and small messages (e.g., votes), respectively. 
This distinction allows us to further refine the timer settings.
The system requires $N\geq 3f+1$, where $f$ is the number of Byzantine faulty nodes.
In a P2P network, messages may be relayed through multiple intermediate nodes before eventually reaching every node.
%
\subsection{P2P networking and libp2p}

Large-scale blockchain systems typically operate over peer-to-peer (P2P) overlay networks that disseminate transactions, blocks, and consensus messages for the consensus layer to establish their ordering and finality.
In P2P networks, several bootstrapping nodes (i.e., bootnodes) are distributed geographically outside the network and provide participation guidance for new nodes.
Although bootnodes are out of the consensus, they help new nodes discover existing peers and obtain basic network metadata.
In a sense, bootnodes serve as a bridge between the system and the outside world.

Traditional Byzantine fault-tolerant (BFT) protocols~\cite{pbft,zyzzyva,hotstuff,sbft} assume a fully connected network, where the totality property is ensured by requiring every node to relay every message to all other nodes, increasing message and communication complexities to $N^2$ and $|m|N^2$, respectively.
Introducing a P2P network and its gossip-based dissemination mechanism reduces the message and communication complexities to $Nd$ and $|m|Nd$, where $d$ denotes the average node degree or fanout.

libp2p~\cite{libp2p} is a modular peer-to-peer networking framework for building decentralized applications.
It provides a publish--subscribe abstraction in which peers subscribe to topics and receive messages disseminated through the P2P overlay, without requiring the publisher to maintain a connection to every subscriber.
\texttt{GossipSub} is libp2p's commonly used PubSub routing protocol.
For each topic, GossipSub maintains a bounded mesh of peers.
The mesh is designed to avoid eager all-to-all forwarding: each peer sends full message payloads only to a small set of mesh neighbors, which bounds per-message fanout while still creating multiple dissemination paths across the overlay.
The mesh size is therefore a key performance parameter: a larger mesh can improve reachability and reduce dissemination delay, but also incurs greater payload redundancy and bandwidth consumption.
To control the mesh size while preserving reachability, 
GossipSub separates eager data forwarding from lazy metadata gossip: besides sending full messages over mesh links, peers periodically advertise message identifiers through \texttt{IHAVE} messages, and a receiver that learns about an unseen message can reply with an \texttt{IWANT} message to request the corresponding payload.
The \texttt{IHAVE}/\texttt{IWANT} exchange therefore serves as a discovery and repair  mechanism: it lets peers detect and fetch missing messages while keeping most redundant traffic at the level of compact message identifiers rather than full payloads.

\begin{figure}[tp]
    \centering
    \includegraphics[scale=0.63]{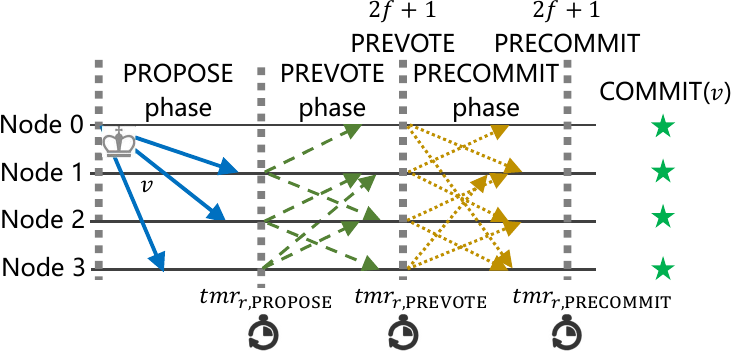}
    \caption{The message pattern of Tendermint ($N=4$).}
    \label{fig:tendermint}
        \vspace{-10pt}
\end{figure}

\subsection{Tendermint and its pipelined extension}
\label{sec:tendermint}

Tendermint~\cite{tendermint} is a leader-based BFT consensus protocol inspired by the seminal PBFT~\cite{pbft}.
Notably, the core idea of Tendermint is to combine the normal-case operations of PBFT with a novel locking mechanism and the underlying P2P network, allowing nodes to reach agreement through a three-phase message pattern even when faulty proposers trigger proposer rotations.
Tendermint also proactively rotates proposers across block heights to ensure a fair distribution of proposal opportunities.

Within each height, Tendermint proceeds through repeated rounds, with each containing \propose, \prevote, and \precommit phases.
The message pattern is depicted in Figure~\ref{fig:tendermint}. 
In each round, a designated proposer or leader broadcasts a proposal, i.e., a candidate block. 
Other nodes first prevote for a valid proposal and then precommit after observing more than two-thirds of prevotes for the same proposal. 
A proposal is committed once it receives precommits from more than two-thirds of the nodes. 
Nodes lock on sufficiently supported proposals to prevent conflicting decisions across rounds, while proposer rotation across rounds ensures progress once the network stabilizes.
Each phase is associated with a timer, upon the expiration of which nodes advance to the next step.

In addition to its locking mechanism, Tendermint relies heavily on the underlying P2P network to eliminate the need for explicit view-change and view-synchronization protocols and dedicated view-change messages, thereby simplifying implementation and maintenance.
More precisely, its liveness depends on the network assumption that, after GST, any message delivered by a correct node is relayed and delivered to all correct nodes within a bounded delay.
Based on this assumption, Tendermint has the following important invariant.

\begin{invariant}
\label{inv:delta}
After GST, all correct nodes advance to the next step within $\Delta$ time.
Here, the next step may correspond to the subsequent phase of the current round, the next round, or even the next height.
\end{invariant}

\noindent\textbf{Pipelined extension.}
Vanilla Tendermint is not pipelined: a proposal at height $h$ is issued only after a proposal at height $h-1$ has been committed.
In the effort to improve our enterprise-grade permissioned blockchain, which uses Tendermint as its consensus module, we found that pipelining is key to boosting the performance of rotating-leader consensus, as also demonstrated by a pipelined HotStuff~\cite{moonshot} or leaderless consensus~\cite{bandle}.
In this work, we target a pipelined variant of Tendermint, where the proposer at height $h$ issues its proposal upon receiving the proposal from height $h-1$.
Nodes may prevote a proposal at height $h$ only after committing a proposal at height $h-1$.

\section{Co-optimization ideas}
\label{sec:ideas}

\subsection{Ledgers for networking}
\label{subsec:network-reconfiguration}

Traditional P2P systems typically manage network membership without relying on centralized coordination. 
For example, SWIM~\cite{swim} decouples failure detection from membership-update dissemination: nodes detect failures through periodic peer-to-peer probing and propagate membership changes by piggybacking them on probe messages. 
In contrast, blockchain systems already provide a replicated state machine through their consensus and contract-execution layers, offering an opportunity to maintain a consistent view of network information and leverage it to optimize the underlying P2P network.

By recording network topology and latency information on-chain, each node maintains a local view of the current network state. 
Upon a membership change, nodes reach consensus on the change to synchronize their views of the updated network information. 
Inspired by the \textit{Group Membership} abstraction~\cite{cachin2011introduction}, we introduce a network configuration abstraction as the core component for managing the underlying network and accommodating network evolution.

\noindent\textbf{Abstraction of network reconfiguration (NR).}
The process by which all nodes obtain consistent information about nodes joining or leaving the network, along with the resulting topology updates, can be abstracted as a \textit{Network Reconfiguration (NR)} primitive.
Assume $\Lambda$ denotes the edge (link) set of the network, i.e., the network topology,
and together with an evolution view identifier $vid$ and the node set $\Pi$
form the network information $\chi = (vid, \Pi, \Lambda)$ kept by every node.
The initial evolution view is 0.
Any node that joins or leaves the network can propose $\chi' = (vid', \Pi', \Lambda')$,
with $vid' = vid + 1$, $\Pi'$ the node set after the change and $\Lambda'$ the consequent network topology,
as an input to the NR.
There may exist more than one $\chi'$ as inputs to the NR.
By consensus, the output of the NR is only one $\chi'$ that every node verifies and installs on its local $\chi$.
The following properties must hold:

(1) Monotonicity: If a correct node installs $\chi = (vid, \Pi, \Lambda)$ and subsequently installs $\chi' = (vid', \Pi', \Lambda')$,
then $vid < vid'$.

(2) Agreement: If some correct node installs $\chi = (vid, \Pi, \Lambda)$ and another correct node installs some $\chi' = (vid, \Pi', \Lambda')$,
then $\Pi = \Pi', \Lambda = \Lambda'$.

(3) Completeness: If a node $p$ (i) joins (ii) leaves / crashes, then eventually every correct node installs $\chi = (vid, \Pi, \Lambda)$ such that (i) $p\in\Pi$ (ii) $p\notin\Pi$.

\begin{figure}[tp]
    \centering
    \includegraphics[width=\linewidth]{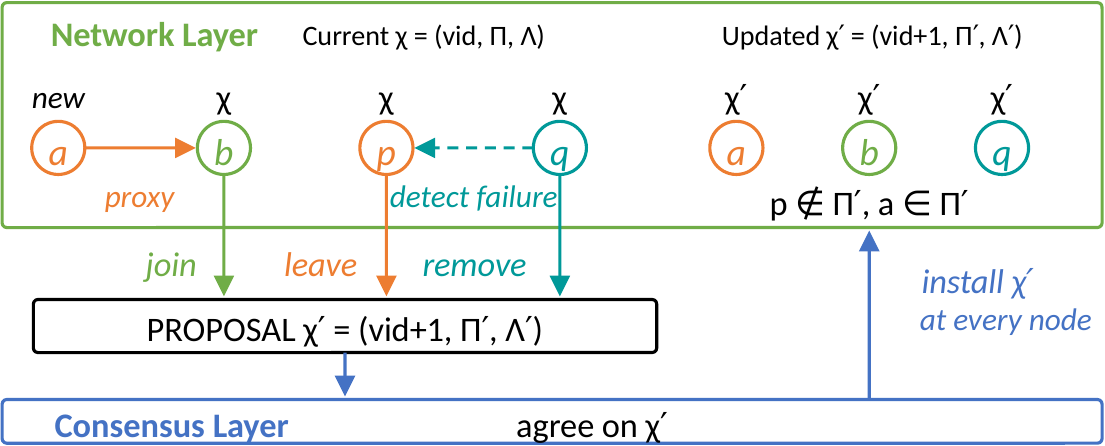}
    \caption{The procedures for network reconfiguration.}
    \label{fig:case}
    \vspace{-10pt}
\end{figure}

Figure \ref{fig:case} depicts the procedures for the network evolution.
The abstraction enables nodes to maintain a consistent view of network information, which can in turn be leveraged to, e.g., organize the network topology more efficiently and/or resiliently~\cite{lowdegree} and improve message routing. 
In this paper, we focus only on the latter problem, while leaving the topology of the underlying system (i.e., libp2p) unchanged.
Note that even if such network information is inaccurate, we can still rely on a fallback mechanism, such as gossip-based dissemination, to ensure reliable message delivery.

In permissionless blockchains such as Ethereum, node operators may be reluctant to expose their complete peer connections. 
Conventional peer-discovery mechanisms~\cite{bitcoinp2p,ethlibp2p} primarily expose network endpoints, whereas revealing peer connections additionally exposes information about the underlying network topology. 
Such information may increase nodes' vulnerability to targeted attacks~\cite{shi2026eclipse,heilman2015eclipse}.
To mitigate this concern, nodes may expose only a subset of their peer connections.
Our experiments in \S\ref{subsec:scaling} and Figure~\ref{fig:scaling-pareto} demonstrate that, even when each node exposes only 50\% of its connections, \sys achieves 90\% of the throughput obtained with full topology visibility under the same configuration.
 
Another concern is the storage overhead of the recorded information, which scales with the size of the network graph, i.e., \(O(N+E)\) for \(N\) nodes and \(E\) connections.
We expect this overhead to remain manageable in the foreseeable future, even as the network scales. 
For example, Ethereum already supports a validator population on the order of one million~\cite{ethvalidator}.

\subsection{Optimized proposer sequence}

The proposer sequence refers to the order in which nodes are selected to propose blocks.
The de facto approach to proposer assignment, particularly in permissioned blockchains, is to use a round-robin scheme~\cite{tendermint,hotstuff}.
In permissionless settings, such a sequence may instead be determined using randomness.
Due to latency disparities among geo-distributed nodes, proposer rotation can incur non-negligible overhead even in a fully connected network, making the proposer sequence an important factor in consensus performance.

Assume there are \(N\) geographically distributed proposers across the globe, with pairwise communication delays denoted by \(d(i,j)\). 
Consider a random path that visits every proposer exactly once and returns to the starting proposer, i.e., it forms a cycle $(v_1,v_2,\dots,v_N,v_1)$.
The total latency of the cycle is $L=\sum_{k=1}^{N} d(v_k,v_{k+1})$, $v_{N+1}=v_1$.
The expected latency of a uniformly random cycle is $\mathbb{E}[L]=N\bar d$,
where $\bar d=\frac{2}{N(N-1)}\sum_{i<j} d(i,j)$
is the average pairwise latency among all proposer pairs.

In practice, intercontinental RTTs between AWS regions can range from 150 to 350 ms~\cite{awslatency}.
Consider a hypothetical proposer distribution and rotation scenario, where proposers are arranged sequentially along the equator and rotate according to their geographical order.
In this case, the total propagation time along the rotation cycle over optical fiber is only a few hundred milliseconds, regardless of \(N\).
We thus have the following observation.

\begin{observation}
The earlier a proposer receives the preceding proposals, the sooner it can generate its own proposal.
\end{observation}

Consider a P2P network where nodes are not fully connected, the problem is formulated as a (variant of) traveling salesman problem~\cite{tsp} (TSP) on a weighted undirected graph.
Given a weighted undirected graph $G=(V,E,w)$, find a cycle such that every vertex in \(V\) is visited exactly once, and the total weight $\sum_{i=1}^{n} w(v_i,v_{i+1})$ ($v_{N+1}=v_1$) is minimized.

\subsection{Efficient and reliable message propagation}
\label{sec:twophasebroadcast}

Byzantine reliable broadcast protocols~\cite{bracha,cachin2011introduction,scalablebrb} commonly rely on all-to-all message propagation to ensure totality, which is essential for the liveness of consensus protocols built upon them (and, in synchronous settings, also for safety~\cite{backbone}).
In the gossip-based communication model, each node instead propagates messages to all or a subset of its neighbors, such that totality is guaranteed with high probability.
Although redundant propagation provides strong robustness, it (unnecessarily) consumes bandwidth in WANs, where bandwidth is both costly and shared among competing messages.

\begin{figure}[tp]
    \centering
    \includegraphics[width=1.00\linewidth]{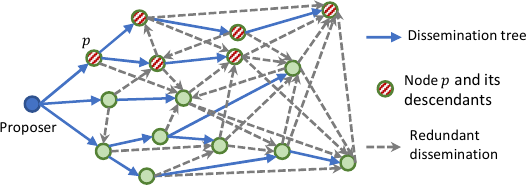}
    \caption{Illustration of proposal dissemination in tree-based and gossip-based Communication Networks.
    Through gossip, each node propagates messages to three neighbors.
    Assume $f=5$ and there are a total of $N=3f+1=16$ nodes.
    Without redundant dissemination, if node $p$ crashes, its descendants in the tree can no longer receive the proposal.
    As they constitute more than $f$ nodes, the upper consensus layer can no longer guarantee liveness, even if the proposer behaves correctly.}
    \label{fig:illustration}
\end{figure}

Following the common principle of consensus protocol optimization that the normal case should be made as efficient as possible, such redundancy can be eliminated~\cite{kauri} as long as messages are delivered successfully and in a timely manner. 
Without redundant propagation, the resulting message dissemination structure becomes a tree.
Assume each node propagates messages to $K$ neighbors.
Gossip-based dissemination incurs $K\times N \times |m|$ communication costs, while such value is $(N-1)\times |m|$ for tree-based dissemination.
However, we must account for cases in which some non-leaf nodes are faulty~\cite{kauri}.
Figure~\ref{fig:illustration} depicts examples of gossip-based and tree-based dissemination.

In fully connected settings, all-to-all flooding may be avoided in normal cases through a pull-based dissemination mechanism~\cite{pullbased}, as leaders or proposers are required to send their proposals to all nodes.
Provided that the proposer is correct and the network remains synchronous, the protocol can proceed regardless of the presence of up to $f$ problematic nodes.
Nodes that do not possess the original proposal can request others to retransmit it.
However, such a pull-based mechanism does not readily apply to P2P networks, as any faulty non-leaf node in a dissemination tree may disrupt reliable message delivery, even when the proposer is correct.

\begin{figure}[tp]
    \centering
    \includegraphics[width=0.95\linewidth]{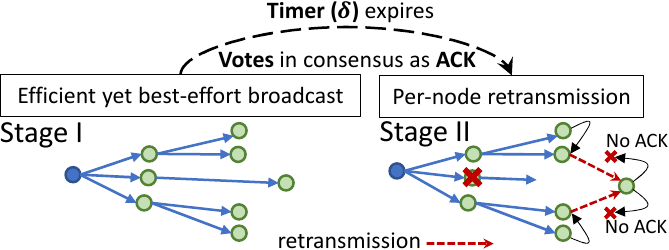}
    \caption{Two-stage reliable broadcast.}
    \label{fig:rbc}
\end{figure}

As blocks or proposals carrying transaction payloads dominate bandwidth consumption, we focus on optimizing proposal dissemination in this paper.
We make the following key observation. 
\begin{observation}
A common pattern in most BFT protocols is for every node to vote on a proposal using a dedicated message, which can naturally serve as an acknowledgment of proposal delivery.
\end{observation}
We thus divide the proposal-dissemination step into two stages: a highly-efficient tree-based broadcast stage and a gossip-based re-transmission stage.
These two stages are linked by the votes cast by individual nodes, with a timer triggering the transition from the first to the second (see Figure~\ref{fig:rbc}).
More interestingly, this idea can be generalized iteratively, with messages from subsequent rounds or heights serving as acknowledgments of the delivery of those from preceding rounds or heights.
We leave this generalization to future work to keep the current design simple and concise.

As proposal dissemination is now split into two stages, we must re-examine the guarantees provided by the upper-layer consensus protocol. 
In particular, \textbf{all properties guaranteed by the original protocol must be preserved}. 
For synchronous protocols~\cite{xft,synchotstuff,themix}, both safety and liveness require careful reconsideration. 
For partially synchronous and asynchronous protocols, only liveness is affected, as safety is ensured by quorum intersection~\cite{quorum}.

In addition to minimizing redundancy and bandwidth consumption, tree construction should consider the propagation delay to future proposers, the maximum end-to-end latency (i.e., the latency to the farthest node), and the maximum node fanout.
Once the topology of $N$ nodes is recorded through consensus, the dissemination tree of each proposer can be determined.

\section{\sys}

\sys is primarily a proposal-dissemination protocol co-designed with libp2p and Tendermint, serving as a concrete instantiation of the ideas discussed in \S\ref{sec:ideas}.
In the following, we elaborate on each of the optimizations introduced above.

\subsection{Greedy algorithm for arranging proposers}
\label{sec:proposer-ring}

Given a weighted undirected graph $G=(V,E,w)$, where vertices represent nodes, edges represent network connections, and each weight denotes the RTT between the corresponding pair of nodes, the goal is to find a cycle that visits every vertex in \(V\) at least once while minimizing the total weight $\sum_{i=1}^{N} w(v_i,v_{i+1})$, where $v_{N+1}=v_1$.

For the complexity analysis, we consider a sparse connection graph $G$ with average node degree $d=O(\log N)$, giving $|E|=O(N\log N)$.
The concrete topology configurations used in our experiments are described in \S\ref{subsec:evaluation-setup}.
We first solve the all-pairs shortest-path problem by running Dijkstra's algorithm $N$ times.
Since the graph is sparse, using a binary heap yields a total time complexity of $O(V(V+E)\log V)=O(N^2\log^2N)$, as $|V|=N$.
We thus reduce the original problem to the Traveling Salesman Problem (TSP), which is still NP-hard.
We employ a simple greedy heuristic to obtain a practical solution by iteratively selecting the unvisited node closest to the current one.
The procedure is depicted in Algorithm~\ref{alg:ring}.
The overall time complexity is dominated by the $N$ executions of Dijkstra's algorithm.

\begin{algorithm}[t]
  \caption{Proposer ring construction.}
  \label{alg:ring}
  \begin{algorithmic}[1]
	\State \textbf{function} ConstructRing($G$) \Comment $G=(V,E,w), |V|=N$
    \State \myindent $M\leftarrow AllPairsDijkstra(G)$
    \State \myindent $p_0\leftarrow Random(V)$
    \State \myindent $U\leftarrow V \setminus \{p_0\}$
    \State \myindent \textbf{for} $i=1..N-1$ \textbf{do}
    \State \myindent\myindent $p_i\leftarrow \argmin_{v\in U} M[p_{i-1}][v]$
    \State \myindent\myindent $U\leftarrow U\setminus \{p_i\}$
	\State \textbf{return} $p_0,p_1,...,p_{N-1}$
  \end{algorithmic}
\end{algorithm}


\subsection{Multi-Factor-Aware Tree Dissemination}
\label{sec:tree}

We aim to construct $N$ cost-efficient dissemination trees based on the network topology recorded on-chain.
We simplify the problem by considering each proposer independently.
The goal is to construct, for each proposer, a spanning tree (of $G$) that minimizes the maximum delay to any node.

If only latency is considered, the optimal solution is a shortest-path tree (or a breadth-first search tree when all latencies are equal) rooted at the proposer.
However, achieving efficient proposal dissemination requires taking several factors into account.
\begin{itemize}
\item Latency: for geo-distributed deployments, proposals must be efficiently disseminated across regions while minimizing unnecessary long-haul transmissions;
\item Bandwidth: given the limited bandwidth available at each node, its fanout must be constrained accordingly;
\item Proposer rotation: to enable rapid proposer handoff, each proposer's dissemination tree should prioritize the next few proposers in the sequence.
\end{itemize}

Because each node's bandwidth is shared among multiple connections, concurrent transmissions inevitably introduce bandwidth contention.
Therefore, the transmission delay of each edge cannot be determined independently, making a shortest-path-tree solution inapplicable.
We further simplify the bandwidth constraints by considering only the fanout of the forwarding node on the last hop, namely node $u$ when trying to add edge $(u,v)$.

The goal is to select $N-1$ edges that connect all nodes
in $G$ while minimizing the maximum delay from the proposer
to any node.
As $N$ may be large, we adopt a simple greedy algorithm that selects $N-1$ edges incrementally.
Initially, only the proposer is included in the dissemination tree, and its outgoing edges are inserted into a binary heap, which maintains the edges with exactly one endpoint in the tree.
The edges are ranked by the propagation delays from the proposer to their endpoints outside the tree.
In each iteration, we extract $\log n$ edges from the heap and select, among them, the edge with the minimum time according to the following formula.
\[
time(u,v)=delay(v)+|m|\times\frac{deg(u)}{B(u,v)},
\]
where $deg(u)$ is the fanout of $u$ after adding $(u,v)$ and $B(u,v)$ is the bandwidth of $(u,v)$.
$delay(v)$ represents the propagation delay from the proposer to $v$, computed as $delay(v)=delay(u)+\mathrm{RTT}(u,v)/2$ for a candidate edge $(u,v)$, and the second term in $time(u,v)$ approximates the transmission delay of message $m$ over $(u,v)$.
The time complexity for constructing $N$ dissemination trees is thus $O(N^2\log^2N)$.
Finally, to prioritize upcoming proposers, when the estimated delays of two edges differ by less than $10\%$, we select the edge whose destination is closer to the current proposer in the proposer sequence.

For simplicity, we estimate the bandwidth of edge $(u,v)$ based on its RTT.
Motivated by the window-limited TCP throughput model~\cite{mathis1997macroscopic}, which predicts that single-flow throughput over a long-haul link is inversely proportional to its RTT, we set
\[
B(u,v)=\min\!\big(B_{\max},\ \kappa/\mathrm{RTT}(u,v)\big),
\]
where $\kappa$ and $B_{\max}$ are configurable parameters adapted to the target deployment and hardware.
We use $|m|$ as a configurable reference payload size for tree construction.

\subsection{Tree-based dissemination with Tendermint}
\label{sec:tree-tendermint}

\begin{figure*}
    \centering
    \subfloat[Normal case]
    {\includegraphics[scale=0.60]{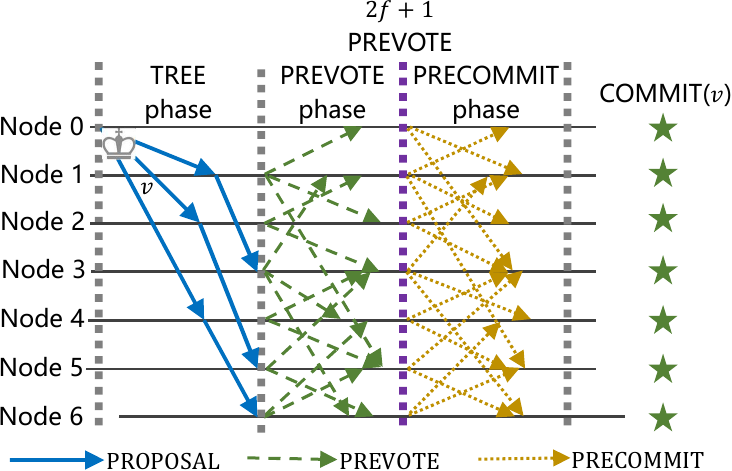}\label{fig:normal}}
    \hspace{15pt}
    \subfloat[Crash case]
    {\includegraphics[scale=0.60]{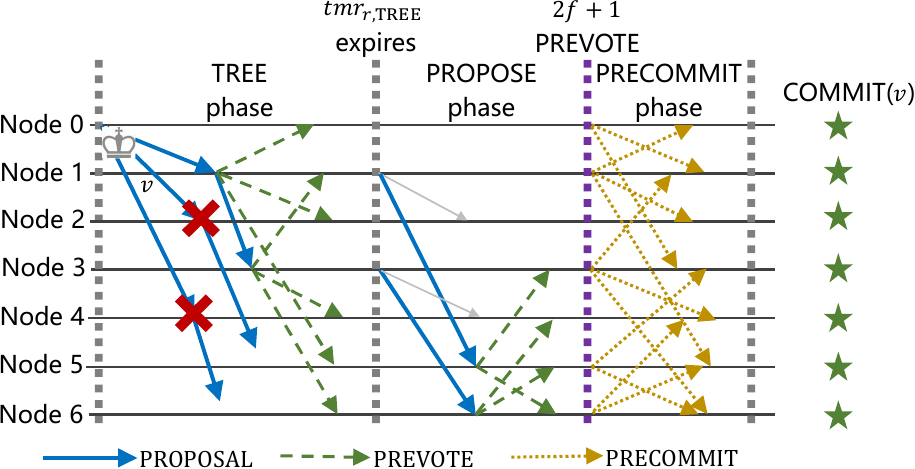}\label{fig:gossip}}
    \caption{Examples for \sys tree-based dissemination ($N=7$).
    In Figure~\subref{fig:normal}, the \propose phase is skipped as all correct nodes receive proposal $v$ during the \best phase.
    In Figure~\subref{fig:gossip}, nodes 2 and 4 crash; consequently, nodes 5 and 6 cannot receive $v$ during the \best phase.
    When $\tmr_{r,\best}$ expires, the neighbors of nodes 5 and 6, relay $v$ to them because neither node has sent a \prevote{} message.}
    \label{fig:example}
    \vspace{-10pt}
\end{figure*}

As we described in \S\ref{sec:tendermint}, Tendermint exchanges three types of messages in each round, corresponding to three phases: \proposal{}, \prevote{} and \precommit{}.
We apply the idea discussed in \S\ref{sec:twophasebroadcast} to \propose{} and \prevote{}, meaning that \proposal{} is disseminated using the tree-based scheme, while \prevote{} implicitly acknowledges the delivery of \proposal{} within the same round.
Both \prevote{} and \precommit{} are still disseminated via gossip.

In this section, we focus primarily on the modifications made to vanilla Tendermint, deferring the full pseudocode to Appendix~\ref{sec:pesudo}.
Since Tendermint broadly relies on timers to drive progress, the two-stage reliable broadcast must be seamlessly integrated with the existing phases and timers without compromising liveness or other properties.
Assume that, via gossip, node $p$ sends messages to the neighbors in $\peer(p)$; via the tree-based dissemination, $p$ instead sends messages to $\children(p)$.
With the two-stage broadcast scheme, we must address two specific issues.

\noindent\textbf{Delivery reliability.}
First, some correct nodes may not be able to receive the proposal through the tree-based dissemination, even if the proposer is correct and the network is synchronous.
To address this issue in Tendermint, sufficient time must be allocated for correct nodes, particularly the proposer, to retransmit the proposal via gossip.

To this end, we explicitly introduce a \emph{tree-based dissemination phase}, denoted by \best{}, along with a corresponding timer, $\tmr_{r,\best}$, at the beginning of each round $r$.
The proposer broadcasts its proposal through tree-based dissemination.
Each node, including the proposer, starts the \best{} timer for height $h$ after receiving $2f+1$ \precommit{}$(v)$ messages at height $h-1$.
Upon receiving a valid proposal $v_r$ for the first time, node $p$ proceeds as in vanilla Tendermint and broadcasts a \prevote{} message for $v_r$.
If node $p$ is still in the \best phase, it also relays the proposal $v_r$ to $\children(p)$, its designated children in the dissemination tree and directly enters the \prevote{} phase, thereby skipping the gossip-based \propose{} phase.
When $\tmr_{r,\best}$ expires, 
node $p$ relays $v_r$ to the nodes in $\peer(p)$ from which it has not received \prevote{} messages for $v_r$, effectively falling back to per-node retransmission.

Otherwise, if timer $\tmr_{r,\best}$ expires before node $p$ receives a valid proposal in round $r$, $p$ proceeds to the \propose phase and starts timer $\tmr_{r,\propose}$, as in vanilla Tendermint.
Upon expiration of $\tmr_{r,\propose}$, node $p$ also follows vanilla Tendermint by broadcasting a \prevote{} $nil$ message and proceeding to the \prevote{} phase.
Figure~\ref{fig:example} illustrates examples of message dissemination in \sys.

With the introduction of the \best{} phase, a correct node that has not received a valid proposal waits for both $\tmr_{r,\best}$ and $\tmr_{r,\propose}$ to expire before sending a $nil$ \prevote.
The \propose{} timeout must account for differences in when nodes enter the round and for proposal delivery through gossip after the \best{} timeout.
We derive the required timer bounds below.

\noindent\textbf{Reorg resilience.}
Introducing two-stage broadcast into Tendermint gives rise to another subtle issue that must be addressed to provide reorg resilience~\cite{moonshot}.
Specifically, \emph{after GST, any valid proposal issued by a correct proposer must be committed, despite the presence of up to $f$ Byzantine nodes}.
This requirement yields to the following key observation.
\begin{observation}
The proposer $p$ at height $h$ must have sufficient time to disseminate its proposal, even if it enters height $h$ later than other correct nodes due to delayed delivery of proposals from preceding heights.
\end{observation}
Otherwise, if proposer $p$ enters \(h\) too late,
the $\tmr_{r,\best}$ and $\tmr_{r,\propose}$ timers of other correct nodes may expire before they receive $p$'s proposal, thereby preventing \(p\)'s proposal from being committed.
This situation in fact violates invariant~\ref{inv:delta}, since $p$ enters height $h$ only after receiving the proposals for all preceding heights.
For instance, assume node $i$ is the first correct node that enters height $h$, doing so at time $t$.
Since node $i$ may advance rapidly without encountering any timeout events, a slower but correct node, particularly the proposer $p$ at height $h$, may still be at height $h-K$, where $K$ can be arbitrarily large.
In case $p$ fails to receive the proposals for heights $h-K$ through $h-1$ via tree-based dissemination, it may take up to $2\delta\times K$ time to catch up to height $h$, as $p$ must obtain these proposals through gossip.
Notably, vanilla Tendermint does not suffer from this issue because all messages, particularly proposals, are disseminated via gossip.

To address this issue, nodes entering a new height must wait sufficiently long before proceeding to the \prevote{} phase. 
That is, the duration of $\tmr_{r,\propose}$ at height $h$ must be long enough to accommodate the cumulative delays introduced by $\tmr_{r,\best}$ at all preceding heights.
A trivial solution is to introduce a timer of duration $2\delta \times K$, but this is impractical.

We observe that this problem arises because gossip dissemination is triggered sequentially, only upon the expiration of $\tmr_{r,\best}$ at each height.
To eliminate these cumulative delays, we modify the protocol to initiate $\tmr_{r,\best}$ for height $h$ proactively, as soon as a node receives $2f+1$ \precommit{} messages for some value $v$ at height $h-1$.
Since \precommit{} messages are disseminated via gossip, all correct nodes can start the corresponding timers within $\Delta$ time (or $\Delta_S$ for small messages) and subsequently switch to gossip dissemination for proposals at both preceding and current heights.

Moreover, the design of the subsequent phases must also be reconsidered.
Under tree-based dissemination, the timer $\tmr_{r,\prevote}$ may expire at some correct nodes before they receive $2f+1$ \prevote{} messages for $v$, causing them to issue nil \precommit{}.
The time gap between the first and last correct nodes to receive the proposal can be as large as $2\Delta+\tmr_{r,\best}$ (or $\Delta_S+\tmr_{r,\best}+\Delta_L$, where $\Delta_L$ bounds the transmission delay of large messages such as proposals), with one $\Delta$ ($\Delta_S$) accounting for the initiation of tree-based dissemination and the other ($\Delta_L$) for the propagation of the proposal via gossip after timeout.
We therefore extend the duration of $\tmr_{r,\prevote}$ to $3\Delta+\tmr_{r,\best}$\footnote{In vanilla Tendermint, $\tmr_{r,\prevote}=2\Delta$ ($\Delta_L+\Delta_S$).} ($\Delta_L+2\Delta_S+\tmr_{r,\best}$) to allow sufficient time for the \prevote{} messages issued by all correct nodes to reach every correct node, where the additional $\Delta$ ($\Delta_S$) accounts for propagation of \prevote{}.

Finally, 
the timer $\tmr_{r,\precommit}$ in Tendermint provides a buffer period, which helps the next proposer to receive $2f+1$ \prevote{} messages for some $v$ (if exist) before advancing to the next round, in case some other node is locked on $v$.
However, doing so would require extending of $\tmr_{r,\precommit}$, thereby further delaying the replacement of a faulty proposer.
Our approach instead ensures that all correct nodes remain synchronized during the \precommit{} phase, with a maximum time skew of $\Delta$ ($\Delta_S$) (i.e., Invariant~\ref{inv:delta}).
Upon receiving any $2f+1$ \prevote{} messages for the same height $h$ and round $r$, node $i$ starts the timer $\tmr_{r,\prevote}$, regardless of whether it is currently in the \prevote{} phase or has even reached height $h$.
The \precommit{} timer starts only after both the \prevote{} timer has expired and $2f+1$ \precommit{} messages for the current height and round have been received.
Finally, we introduce an optimization that enables both Tendermint and \sys to advance quickly to a new round.
Upon receiving $2f+1$ nil \prevote{} messages for the current height and round, the node immediately advances to the next round.



The settings of $\tmr_{r,\propose}$, $\tmr_{r,\prevote}$ and $\tmr_{r,\precommit}$ are key to ensuring liveness and are therefore closely tied to the assumed maximum network delay $\Delta$.
In contrast, $\tmr_{r,\best}$ can be set more aggressively to better reflect the actual network delay.
However, setting $\tmr_{r,\best}$  too small may lead to frequent retransmissions, whereas setting it too large delays round advancement upon crash.
We assume $\tmr_{r,\best}$ is set to $2\delta$ and $\tmr_{r,\propose}$ is set to $3\Delta$ ($2\Delta_L+\Delta_S$), 
where $\delta<\Delta$.
Compared with vanilla Tendermint, \sys incurs an additional per-round delay of $\tmr_{r,\best}+\Delta$ ($\tmr_{r,\best}+\Delta_S$) to accommodate scenarios in which correct nodes must advance to the next round following a proposer crash.
Finally, we also apply a simple optimization to both Tendermint and \sys.
Upon receiving $2f+1$ \prevote{} $nil$ in round $r$, a node can immediately advance to round $r+1$, skipping the \precommit{} phase.
This is safe because no value can have been locked by any node or committed in round $r$.

\subsection{Correctness proof}

\sys preserves Tendermint's voting rules, locking mechanism, and quorum requirements, which ensure safety regardless of message delays and timeout durations.
Pipelining only disseminates proposals in advance; correct nodes prevote only after committing the preceding height and validating the proposal against the committed prefix.
Thus, \sys inherits Tendermint's safety.
We therefore focus on proving that, after GST, all correct nodes receive and cast both prevotes and precommits for every proposal issued by a correct proposer, from which both liveness and reorg resilience follow.
Since round advancement is simpler and can be analyzed analogously, we mainly focus on the case where nodes enter a new height.

\noindent\textbf{\prevote{}($v$).}
Suppose node $i$ is the first correct node to initiate $\tmr_{r,\best}$ at height $h$, doing so at time $t$ upon receiving $2f+1$ \precommit{} messages for some value $v$ at height $h-1$.
Thus, every correct node initiates $\tmr_{r,\best}$ by $t+\Delta$ ($t+\Delta_S$).
All correct nodes switch to gossip by time $t+\Delta+2\delta$ ($t+\Delta_S+2\delta$).
Let node $p$ be the proposer of height $h$. 
Node $p$ issues its proposal by time $t+2\Delta+2\delta$ ($t+\Delta_S+\Delta_L+2\delta$), since all preceding proposals can reach $p$ by then.
The proposal then reaches every correct node within an additional $\Delta$ ($\Delta_L$), i.e., at $t+3\Delta+2\delta$ ($t+\Delta_S+2\Delta_L+2\delta$). 
By setting $\tmr_{r,\propose}\ge 3\Delta$ ($\Delta_S+2\Delta_L$), every correct node can receive the proposal before its $\tmr_{r,\propose}$ expires and thus prevotes the proposal.

\noindent\textbf{\precommit{}($v$).}
Each node starts the timer $\tmr_{r,\prevote}$ upon receiving any $2f+1$ \prevote{} messages.
Assume that the first node starts $\tmr_{r,\prevote}$ at time $t$, which could theoretically occur at the very beginning of a round, when the proposal is issued.
Thus, the last correct node may send its \prevote{} as late as $t+2\Delta+2\delta$ ($t+\Delta_S+\Delta_L+2\delta$), after which the message may take up to an additional $\Delta$ ($\Delta_S$) to reach every node.
Thus, setting $\tmr_{r,\prevote}$ to $3\Delta+2\delta$ ($\Delta_L+2\Delta_S+2\delta$) gives all correct nodes until $t+3\Delta+2\delta$ ($t+\Delta_L+2\Delta_S+2\delta$) to issue \precommit{} messages for $v$.


\noindent\textbf{Livelock prevention.}
Finally, we show that setting $\tmr_{r,\precommit}$ to $2\Delta$ ($2\Delta_S$) prevents livelock, where $\tmr_{r,\prevote}=3\Delta+2\delta$ ($2\Delta_S+2\delta+\Delta_L$).
Suppose some correct node, say node $p$, locks on $v$ in round $r$ after GST, and the first correct node starts $\tmr_{r,\prevote}$ at time $t$.
The triggering $2f+1$ \prevote{} messages reach all correct nodes within $\Delta_S$, so every correct node's $\tmr_{r,\prevote}$ expires before $t+\tmr_{r,\prevote}+\Delta_S$.
Since $p$ locks on $v$ only before its $\tmr_{r,\prevote}$ expires, the $2f+1$ \prevote{} messages for $v$ reach all correct nodes before $t+\tmr_{r,\prevote}+2\Delta_S$, and proposal $v$, which some correct node already holds at time $t$ (a lock leaves at most $f$ correct nodes prevoting for a different value), arrives even earlier, by $t+\Delta_S+2\delta+\Delta_L$.
Meanwhile, no correct node leaves round $r$ before $t+\tmr_{r,\prevote}+\tmr_{r,\precommit}$: its $\tmr_{r,\precommit}$ starts only after its own $\tmr_{r,\prevote}$ expires, $2f+1$ nil \prevote{} messages cannot coexist with $2f+1$ \prevote{} messages for $v$, and skipping to a higher round requires some correct node to have left first.
As $t+\tmr_{r,\prevote}+2\Delta_S\leq t+\tmr_{r,\prevote}+\tmr_{r,\precommit}$, every correct node updates $validValue$ and $validRound$ before leaving round $r$, so the next correct proposer's value can be accepted by all correct nodes, and that round decides.

\begingroup
\providecommand{\evalfigdir}{figures}
\raggedbottom
\makeatletter
\@ifclassloaded{acmart}{%
\patchcmd{\subsection}
  {-.75\baselineskip \@plus -2\p@ \@minus -.2\p@}
  {-.5\baselineskip}
  {}{\PackageError{evaluation}{Cannot adjust subsection spacing}{}}
}{}
\makeatother

\section{Evaluation}

\subsection{Implementation and Experimental Setup}\label{subsec:evaluation-setup}

\noindent\textbf{Implementation.}
We implemented \sys on top of Tendermint~\cite{tendermint} and libp2p~\cite{libp2p}.
The code is available at \url{https:
//github.com/Decentralized-Computing-Lab/Duet}.
Specifically, \proposal{} messages are disseminated along each proposer's dissemination tree, while \prevote{} and \precommit{} messages use libp2p's \texttt{gossipsub} with target mesh degree set to six.
For each height and round, nodes record the mesh peers that sent \prevote{}.
When the \best-phase timer $\tmr_{r,\best}$ expires, the retransmission stage pushes the \proposal{} only to those mesh peers from which no \prevote{} has been observed, and sends the corresponding \texttt{IHAVE} messages to its non-mesh peers.

\noindent\textbf{Baselines.}
We compare \sys against two baselines implemented in the same prototype.
\texttt{Gossip} disseminates each \proposal{} over the same \texttt{gossipsub} mesh, whereas \texttt{K-ary tree} disseminates each \proposal{} along a per-proposer balanced tree with branching factor \(K\).
The tree is constructed level by level, with each parent selecting up to \(K\) of its neighbors in \(G\) that have not yet been added to the tree as its children.
For both protocols, \prevote{} and \precommit{} messages use the same \texttt{gossipsub} configuration as in \sys; only the dissemination of \proposal{} messages differs.
We choose these two schemes as baselines because Gossip is the default dissemination mode in libp2p, while the K-ary tree captures the balanced-tree dissemination structure used by Kauri~\cite{kauri}, a state-of-the-art tree-based BFT protocol.

To ensure a fair comparison, we evaluate all protocols over the same network topology $G=(V,E,w)$, such that they differ only in how proposals are disseminated over $G$.
The topology is fully connected for $N=10$, with an average node degree of 20 for $N=50$ and 40 for $N\in\{100,300\}$.
The edge weights correspond to the RTTs measured between the AWS regions in our deployment.
Given $G$, each \sys node deterministically derives the proposer sequence (\S\ref{sec:proposer-ring}) and the corresponding dissemination trees (\S\ref{sec:tree}), forwarding each proposal to its children in the tree associated with the corresponding proposer.
Since these structures are deterministic functions of \(G\), all nodes independently derive identical structures without additional coordination.
For tree construction, we instantiate the bandwidth model described in \S\ref{sec:tree} with $\kappa=20000$ and $B_{\max}=800$\,Mb/s.
We fix $|m|=10$\,MB as a tree-construction parameter across all experiments, while allowing the actual proposal size to vary with the batch size.


\noindent\textbf{Experimental setup.}
Unless otherwise specified, experiments are conducted across 10 AWS regions: three in the US (N.\ Virginia, Ohio, and N.\ California), three in Europe (Ireland, London, and Frankfurt), and four in the Asia-Pacific region (Tokyo, Singapore, Sydney, and Mumbai).
Each node runs on an \texttt{m4.xlarge} instance with 4 vCPUs and 16\,GiB of memory.
By default, nodes are evenly distributed across the regions.
Latency is measured from the time a \proposal{} is issued until it is committed, and each transaction is 1\,KB in size.
Guided by the parameter choices in CometBFT~\cite{cometbft}, we set $\delta=1.5$~s, $\Delta_S=0.5$~s, and $\Delta_L=2.5$~s,  unless otherwise specified.

\subsection{Normal-case performance}
\label{subsec:scaling}

\IfFileExists{\evalfigdir/cross_region_traffic.pdf}{%
\begin{figure*}[t]
  \centering
  \begin{minipage}[t]{0.70\textwidth}
    \centering
    \vspace{0pt}
    \includegraphics[width=\linewidth]{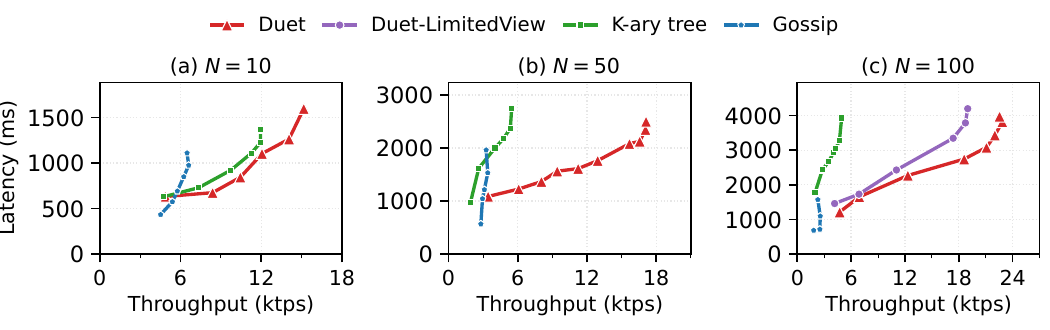}
    \caption{Performance as the system scales ($N{=}10$, $50$, and $100$).}
    \Description{Three performance plots for 10, 50, and 100 nodes.}
    \label{fig:scaling-pareto}
  \end{minipage}\hfill
  \begin{minipage}[t]{0.28\textwidth}
    \centering
    \vspace{0pt}
    \includegraphics[width=\linewidth]{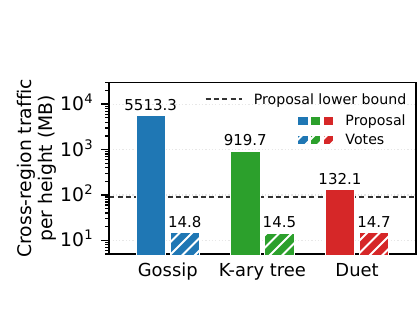}
    \caption{Cross-region traffic.}
    \Description{Proposal and vote traffic for Gossip, K-ary tree, and \sys, with a proposal lower bound.}
    \label{fig:cross-region-traffic}
  \end{minipage}
\end{figure*}
}{%
\begin{figure*}[t]
  \centering
  \includegraphics[scale=0.82]{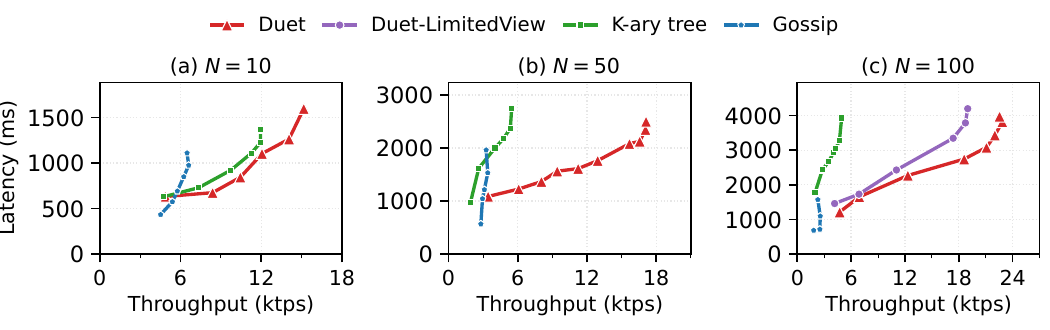}
  \caption{Performance as the system scales from $N{=}10$ to $N{=}100$ (left to right: $N{=}10$, $50$, and $100$).}
  \Description{Three side-by-side performance plots across system sizes from $N=10$ to $N=100$, with a single shared legend at the top, comparing \sys, \sys-LimitedView, Gossip, and a K-ary tree.}
  \label{fig:scaling-pareto}
\end{figure*}
}

\begin{figure}[t]
  \centering
  \includegraphics[scale=0.53]{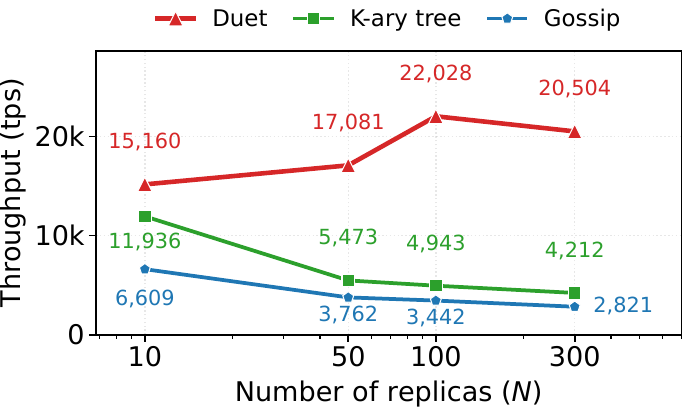}
  \caption{Peak throughput as $N$ scales from 10 to 300.}
  \Description{Line chart of throughput versus number of nodes on a log x-axis.
  \sys rises from about 15k tps at N=10 to a peak near 22k tps at N=100 and holds about 20k tps at N=300.
  K-ary tree falls from 12k tps at N=10 to about 4.2k tps at N=300.
  Gossip falls from 6.6k tps at N=10 to about 2.8k tps at N=300.}
  \label{fig:scaling-throughput-vs-n}
\end{figure}

This experiment evaluates the normal-case scalability of \sys and the other protocols. 
We vary the number of nodes from \(N=10\) to \(N=300\) in the balanced 10-region WAN deployment and gradually increase the batch size until each protocol reaches saturation. 
We set the branching factor of the K-ary tree to \(K=3\) for \(N=10\) and \(K=5\) for \(N\in\{50,100,300\}\).

\begin{figure*}[t]
  \centering
  \includegraphics[scale=0.80]{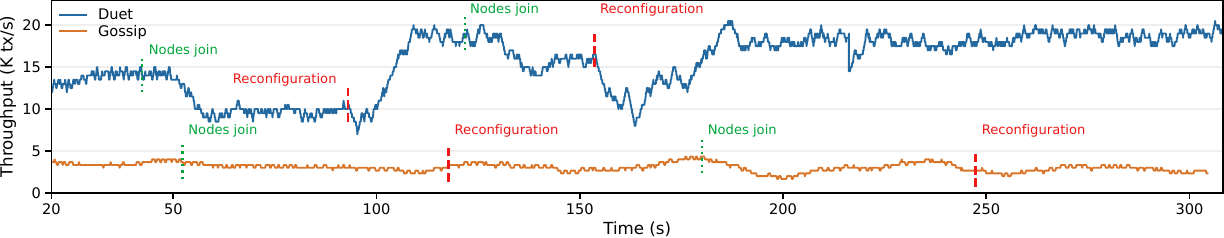}
  \caption{Throughput during dynamic expansion from 10 to 30 and to 50 nodes.}
  \Description{Throughput over time for \sys and Gossip, with two network-joining events and two consensus-reconfiguration events marked separately for each protocol. \sys experiences temporary throughput reductions during joining and reconfiguration, then recovers to a higher steady throughput than in the initial configuration. Gossip maintains lower throughput with a slight overall decline as membership expands.}
  \label{fig:dynamic-membership}
\end{figure*}

As Figure~\ref{fig:scaling-pareto} shows, at $N=10$, Gossip is limited by redundant cross-region forwarding.
With one node per region, mesh-based dissemination repeatedly consumes scarce WAN bandwidth.
The K-ary tree performs considerably better at this scale.
With $K=3$, the K-ary tree has limited depth, allowing upcoming proposers to receive the prefix sufficiently early for the pipeline to utilize the available bandwidth.
Hence, \sys achieves only a modest throughput advantage over the K-ary tree at \(N=10\).

The performance gap widens as the system scales to $N=50$ and $N=100$.
By better accounting for geographic locality, \sys reduces the distance traveled during proposal dissemination.
In contrast, the K-ary tree randomly selects children from each node's remaining neighbors.
Moreover, the K-ary tree is constrained by prefix delivery:
balanced trees may place later proposers behind additional inter-region hops without prioritizing upcoming proposers, delaying prefix delivery and preventing subsequent heights from promptly entering the pipeline.
As a result, the \proposal{} pipeline cannot fully utilize the available bandwidth, causing throughput to drop sharply relative to the 10-node case.

Gossip similarly suffers from delayed prefix delivery, while its fixed target mesh degree introduces additional redundant WAN traffic at every scale.
For both baselines, the additional performance drop from $N=50$ to $N=100$ is smaller because the pipeline is already partially underutilized at $N=50$.
Increasing the network size primarily exacerbates the same bottleneck rather than introducing a new one.

\sys exhibits the opposite trend as \(N\) increases from 10 to 100. 
At larger scales, tree construction benefits from a richer set of low-RTT candidate links for connecting nearby nodes and upcoming proposers. 
This accelerates prefix delivery across consecutive heights, enabling deeper pipelining and more efficient bandwidth utilization.

Figure~\ref{fig:scaling-pareto} also includes \emph{\sys-LimitedView} at \(N=100\), which uses the same underlying topology but allows each node to expose only 20 links (50\%) for tree construction. 
This limited visibility reduces the quality of the resulting trees: \sys-LimitedView achieves 19.00k tps at a latency of 4196\,ms, compared with 21.09k tps at 2899\,ms for \sys.
These results show that greater link visibility enables the construction of more efficient trees, while \sys-LimitedView still substantially outperforms the baselines.

To explain the throughput gap at \(N=100\), we examine the pipeline depth with a batch size of 10k. 
If a proposer has issued a proposal at height \(h_1\) while the latest committed height is \(h_2\), we define the pipeline depth as \(h_1-h_2\). 
\sys achieves an average pipeline depth of 8, compared with 2 for the K-ary tree and 1 for Gossip. 

\IfFileExists{\evalfigdir/cross_region_traffic.pdf}{%
Under the same setting with a batch size of 10k, Figure~\ref{fig:cross-region-traffic} shows the total \emph{cross-region} traffic per height, aggregated across all nodes.
\sys reduces total cross-region proposal and vote traffic by 84.3\% and 97.3\% compared with the K-ary tree and Gossip, respectively.
These reductions stem from two mechanisms that optimize proposal dissemination.
First, tree-based dissemination eliminates redundant forwarding in the gossip mesh: Gossip generates $6.0\times$ as much cross-region proposal traffic as the K-ary tree. 
Second, \sys's tree construction (\S\ref{sec:tree}) favors low-RTT links, accelerating proposal dissemination while exploiting geographic locality to reduce cross-region transfers.
In contrast, the K-ary tree selects children randomly, resulting in cross-region proposal traffic equivalent to approximately 92 full proposals per height, compared with only 13 for \sys---close to the theoretical minimum of nine required to reach the other nine regions.
\par
}{}


Figure~\ref{fig:scaling-throughput-vs-n} extends the scaling experiment to \(N=300\). 
\sys sustains 20.5k tps at \(N=300\), close to its peak throughput of 22.0k tps at \(N=100\), whereas the K-ary tree and Gossip drop to 4.2k and 2.8k tps, respectively. 
At this scale, \sys thus achieves \(7.26\times\) the throughput of Gossip. 
The modest throughput decline of \sys from \(N=100\) to \(N=300\) likely results from diminishing benefits of deeper pipelining and increased gossip-based vote traffic as the network grows.


\begin{figure*}[t]
  \centering
  \begin{minipage}[t]{0.48\textwidth}
    \centering
    \includegraphics[scale=0.60]{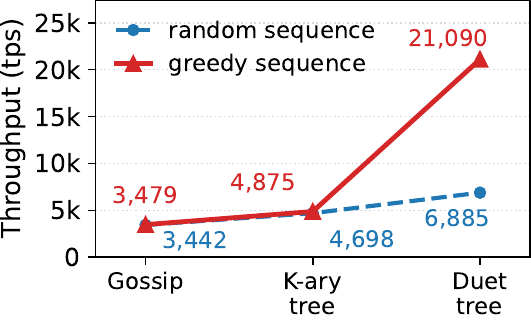}
    \caption{Ablation Study of dissemination and proposer sequence ($N{=}100$, 10-region WAN, batch size 10k).}
    \Description{A line chart plotting throughput against the propagation method (Gossip, K-ary tree, \sys tree) with two lines for the random and greedy proposer sequences.
    Both lines nearly coincide for Gossip and K-ary and then fan apart sharply at \sys tree, where the greedy-sequence line reaches 21,090 tps.}
    \label{fig:ablation-protocol-ring}
  \end{minipage}\hfill
  \begin{minipage}[t]{0.42\textwidth}
    \centering
    \includegraphics[scale=0.43]{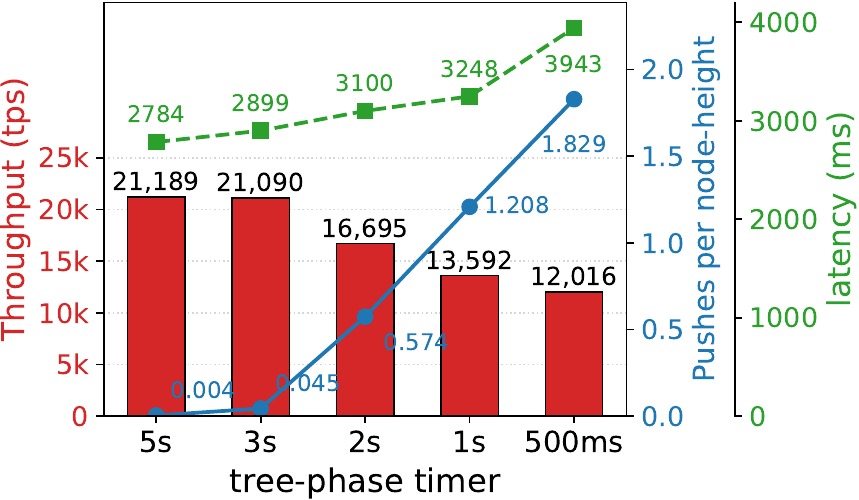}
    \caption{\best{} phase timer sensitivity ($N{=}100$, batch size 10k, fault-free).}
    \Description{A grouped chart with throughput bars, a push-peers-per-node-height line, and an average commit latency line, all plotted against five \best{} phase timer values from 5 seconds down to 500 milliseconds.}
    \label{fig:tree-phase-timer-sweep}
  \end{minipage}
\end{figure*}

\subsection{Dynamic network expansion}\label{subsec:dynamic-membership}

We evaluate the network reconfiguration mechanism described in \S\ref{subsec:network-reconfiguration} by measuring the instantaneous throughput of \sys and Gossip during dynamic network expansion. 
The number of nodes grows from 10 to 30 and then to 50, with the initial 10 nodes and each batch of 20 joining nodes evenly distributed across the 10 regions.
We divide each expansion into two stages, \emph{node joining} and \emph{reconfiguration}, to distinguish the costs and benefits of P2P-layer operations from that of consensus-layer updates.

During node joining, new nodes relay messages but do not participate in consensus voting. 
Both protocols randomly re-select their mesh peers after the new nodes join. 
During reconfiguration, both protocols reach consensus on the updated node set, while \sys additionally reach consensus on the peer connections and RTTs of the new configuration.
After reconfiguration, the new nodes begin voting in both protocols, while \sys also switches to the newly derived proposer ring and dissemination trees.

Figure~\ref{fig:dynamic-membership} shows that the throughput of \sys declines after new nodes join,
as new nodes cannot provide acknowledgments through their own \prevote{} messages until they become voting members. 
They however remain eligible for retransmission, introducing additional traffic that competes for bandwidth. 
During reconfiguration, \sys switches to the newly derived proposer ring and dissemination trees and reestablishes its pipeline, resulting in a brief throughput drop followed by recovery.
In contrast, Gossip sustains the same dissemination scheme across reconfiguration, so the transition itself has only a limited effect. 
However, the larger consensus group increases dissemination overhead, reducing its steady-state throughput.


\subsection{Ablation study}\label{subsec:ablation}

To isolate the effects of proposal dissemination and proposer sequence on performance, we combine each of three dissemination schemes \{Gossip, K-ary tree, \sys{}\} with each of two proposer sequences \{random, greedy\}.
Gossip uses the same unstructured mesh as in \S\ref{subsec:evaluation-setup}, while the K-ary tree uses balanced per-proposer trees with \(K{=}5\). 
With random proposer sequence, the dissemination tree of \sys is still constructed according to the rule described in \S\ref{sec:tree}. 


Figure~\ref{fig:ablation-protocol-ring} shows that Gossip is  insensitive to proposer sequence.
Its mesh is unstructured: a \proposal{} reaches the next proposer through multi-hop mesh dissemination and redundant forwarding, rather than along paths optimized for geographic locality. 
As a result, replacing the random sequence with the greedy sequence has little effect on throughput.
The K-ary tree exhibits the same limitation despite eliminating redundant dissemination, as its balanced trees are constructed randomly. 
A greedy sequence does not ensure that the next proposer is nearby in the dissemination tree, so subsequent heights must wait for prefix delivery before entering the pipeline. 


\sys benefits most from combining its dissemination trees with the greedy proposer sequence. 
The dissemination trees prioritize prefix delivery to upcoming proposers, while the greedy sequence places consecutive proposers closer to one another, allowing proposals to reach their successors more quickly. 
Together, these mechanisms accelerate prefix delivery across consecutive heights, enabling deeper pipelining and substantially higher throughput. 


\subsection{Timer Sensitivity}
\label{subsec:tree-phase-timer}

The gossip-based retransmission stage is highly sensitive to the duration of the \best{}-phase timer. 
A shorter timer triggers retransmission more frequently because it is more likely to expire during normal execution before the corresponding \prevote{} messages arrive.
Figure~\ref{fig:tree-phase-timer-sweep} studies this effect in a  100-node, batch-size-10k setting without faults.
The ``pushes per node-height'' metric measures the average number of mesh peers, among the 6 peers of each node, to which the node pushes the \proposal{} for retransmission at each height.


With a 5\,s or 3\,s \best{}-phase timer, retransmission is rarely triggered under fault-free conditions. 
At 2\,s, some \prevote{} messages arrive after the timer expires, causing nodes to initiate retransmission even though tree-based dissemination would otherwise complete successfully. 
This effect becomes more pronounced at 1\,s and 500\,ms, with retransmission traffic exceeding one push per node-height.
The \best{}-phase timer is therefore a critical deployment-specific parameter that must be carefully tuned to balance timely recovery against unnecessary retransmission.


\subsection{Ethereum-Like heterogeneous deployment}
\label{subsec:hetero}

\begin{figure}[t]
  \centering
  \includegraphics[scale=0.60]{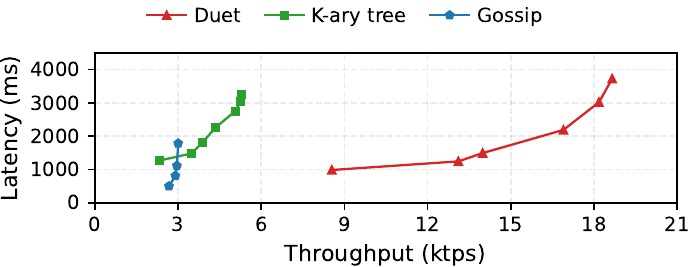}
  \caption{Throughput--latency under the Ethereum-like deployment ($N{=}100$).}
  \label{fig:pareto-heterogeneous}
\end{figure}

\begin{figure*}[!htbp]
  \begin{minipage}[t]{0.45\textwidth}
    \centering
    \includegraphics[scale=0.58]{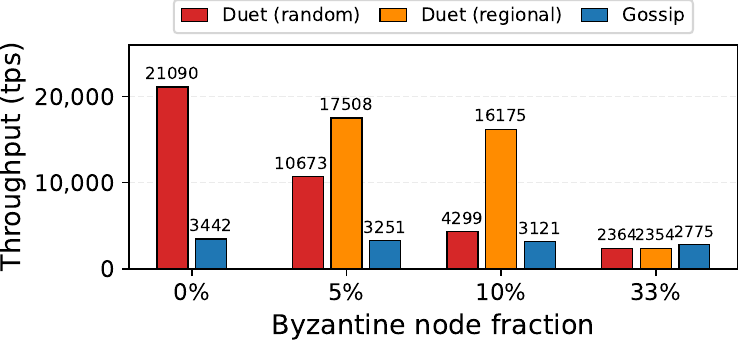}
    \captionof{figure}{Throughput with Byzantine non-leader nodes.}
    \Description{Grouped bar chart of throughput at 5\%, 10\%, and 33\% Byzantine nodes: \sys with random placement (red), \sys with region-concentrated placement (orange), and Gossip (blue).
    At 5\% and 10\% Byzantine nodes, region-concentrated placement retains the most throughput (17.5k then 16.2k) and random placement drops sharply (10.7k then 4.3k), both well above Gossip ($\sim$3.1--3.3k).
    At 33\% Byzantine nodes all three drop to roughly 2.3--2.8k, with Gossip marginally highest.}
    \label{fig:crash-sweep}
  \end{minipage}\hfill
  \begin{minipage}[t]{0.45\textwidth}
    \centering
    \includegraphics[scale=0.56]{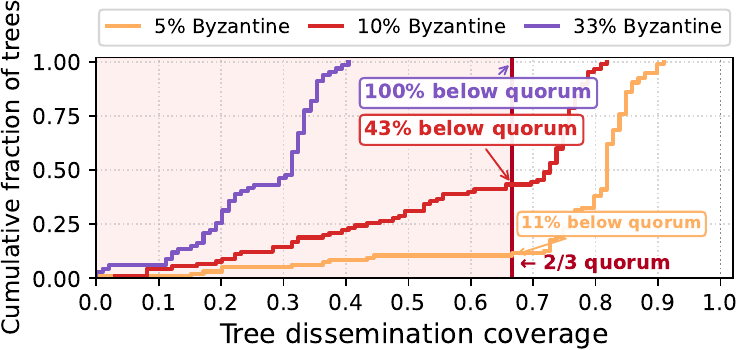}
    \captionof{figure}{CDF of per-tree dissemination coverage under randomly distributed Byzantine nodes.}
    \Description{Cumulative distribution function plot with three step lines for 5\%, 10\%, and 33\% randomly placed Byzantine nodes.
    The 5\% curve has a long tail crossing into the below-quorum region (11\% below the 2/3 threshold); the 10\% curve crosses the threshold at 43\%; the 33\% curve is entirely below the threshold.}
    \label{fig:crash-quorum-cdf}
  \end{minipage}
\end{figure*}

\begin{figure}[!htbp]
  \centering
  \includegraphics[scale=0.86]{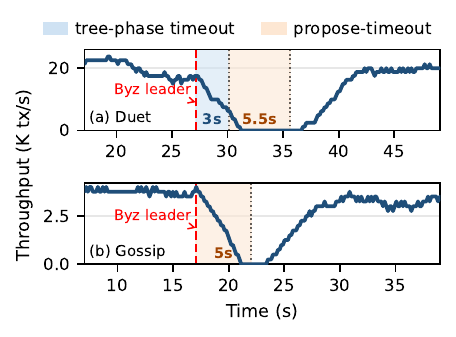}
  \caption{Throughput when a Byzantine leader withholds its proposal.}
  \label{fig:leader-failure-timeline}
\end{figure}

To evaluate \sys under a more realistic deployment, we construct an Ethereum-like node placement based on the country-level distribution of nodes reported by Ethernodes~\cite{ethernodes}. 
We map each country to its nearest AWS region and normalize the resulting distribution to 100 nodes. 
The resulting placement is concentrated in Frankfurt (30 nodes) and N.\ Virginia (20 nodes), with 39\% of the nodes in North America, 46\% in Europe, and 15\% in the Asia-Pacific region. 
The complete per-region distribution is provided in Appendix~\ref{app:placement}.


Figure~\ref{fig:pareto-heterogeneous} shows that \sys still retains a clear performance advantage under this skewed placement. 
Gossip and the K-ary tree do not explicitly exploit the regional skew: Gossip disseminates proposals over an unstructured mesh, while the K-ary tree constructs balanced trees without considering geographic locality.
In contrast, \sys benefits from the regional skew by constructing multi-factor-aware trees that prioritize low-latency paths within the node-dense regions of Europe and the US East. 
Comparing the two deployments at a throughput of approximately 13--14k tps in Figure~\ref{fig:scaling-pareto} (\(N=100\)) and Figure~\ref{fig:pareto-heterogeneous}, \sys achieves a latency of 1493\,ms under the skewed placement, compared with 2269\,ms under the uniform placement, benefiting from the greater geographic locality of nodes in the skewed deployment.


\subsection{Byzantine Behaviors and Limitations}
\label{sec:resilience}

\noindent\textbf{Byzantine Non-Leader Nodes.}
We vary the fraction of Byzantine nodes (5\%, 10\%, and 33\%) that remain online but deliberately refuse to forward \proposal{} and \prevote{} messages. 
We exclude these nodes from the proposer sequence to isolate their impact on message dissemination and consider two Byzantine-node placements: \emph{random} and \emph{region-concentrated}. 
At 33\%, the region-concentrated placement covers all nodes in three regions and three additional nodes in a fourth region.

Figures~\ref{fig:crash-sweep} and~\ref{fig:crash-quorum-cdf} show that Byzantine nodes degrade \sys's performance by disrupting multiple trees.
With 5\% and 10\% Byzantine nodes, respectively, 11\% and 43\% of the trees fail to reach a quorum through the tree-based dissemination alone. 
As a result, subsequent proposers must wait for retransmission to deliver the missing prefixes after the 3\,s \best{}-phase timer expires.
When Byzantine nodes are geographically concentrated, \sys's RTT-aware routing places fewer correct nodes behind Byzantine nodes in the dissemination tree, allowing tree-based dissemination to reach a larger fraction of correct nodes.
At 33\%, both placements rely heavily on retransmission, and the additional waiting time introduced by the \best{}-phase timer reduces \sys's throughput below that of Gossip.


\noindent\textbf{Byzantine Leaders.}
We next consider a Byzantine leader that deliberately refrains from proposing a block, leaving no \proposal{} available for retransmission. 
Compared with Gossip, \sys waits for an additional \best{} phase and uses a longer \propose{} timeout (\S\ref{sec:tree-tendermint}).
Consequently, \sys takes longer to resume progress but recovers its previous throughput once a correct leader takes over (Figure~\ref{fig:leader-failure-timeline}). 
Across both Byzantine experiments, the 3\,s \best{} timer strikes a balance between recovery speed and normal-case throughput, as shorter timers accelerate recovery but may trigger unnecessary retransmissions during normal operation (\S\ref{subsec:tree-phase-timer}).


\par
\endgroup

\section{Related Work}

\noindent\textbf{P2P networks.}
P2P overlay networks have been extensively studied since the early 2000s~\cite{chord,ODRI,Pastry,CAN,Viceroy,survey}.
Most previous work focused on structured overlays, where a well-defined (and rather intricate) rule is used to guide node connections and routing.
It would be interesting to further extend the ideas of \sys to structured overlays.
More recently, Aradhya et al.\cite{overlaysmeetblockchain} also proposed a cross-layer design in which the blockchain assists in maintaining its underlying P2P overlay and enables recovery from catastrophic failures. 
In contrast, we simply record network topologies and latencies via the blockchain and focus on proposer rotation and message propagation.
Prior P2P systems have extensively explored decentralized network-location services.
For example, Meridian~\cite{meridian} organizes peers into multi-resolution rings based on direct latency measurements and uses gossip to maintain these local views, enabling scalable latency-aware node selection without constructing global network coordinates. 
In contrast, \sys maintains a globally consistent view of network conditions and exploits an upper-layer consensus protocol to coordinate network information and optimize protocol-level message dissemination.

\noindent\textbf{Tree-based BFT dissemination and aggregation.}
Several BFT protocols organize communication using trees. 
ByzCoin~\cite{byzcoin} uses communication trees and collective signing to improve the scalability of Byzantine consensus. 
Kauri~\cite{kauri} introduces pipelined tree-based dissemination and aggregation for HotStuff-style protocol. 
Our dissemination tree follows a design similar to those of ByzCoin and Kauri, but \sys adopts a complementary recovery mechanism tailored to P2P overlays. 
Rather than reconfiguring the tree or falling back to one-to-all communication, \sys resorts only to per-node retransmission via gossip, triggered by the absence of \prevote{} messages serving as acknowledgments.
Moreover, \sys optimizes its dissemination trees by jointly considering network topology, link latency, node fanout, and the positions of upcoming proposers.

\noindent\textbf{Network-layer optimization.}
Some work optimizes message dissemination without fundamentally modifying the upper-layer consensus protocol.
Graphene~\cite{graphene} compresses block propagation through interactive set reconciliation, while Erlay~\cite{erlay} reduces Bitcoin transaction-relay overhead by replacing extensive flooding with efficient reconciliation. 
FRING~\cite{fring} constructs a geography-aware P2P overlay for blockchain systems and introduces a broadcast algorithm that reduces redundant transmissions while preserving sufficient robustness. 
These approaches optimize the network layer but generally do not exploit consensus semantics. 
In contrast, \sys leverages acknowledgments provided by the consensus layer to eliminate redundant transmissions.

\noindent\textbf{Network-aware consensus.}
A broader body of work exploits network structure or locality to accelerate replication. 
Ring Paxos~\cite{ringpaxos} organizes communication in a ring to achieve high-throughput atomic broadcast. 
WPaxos~\cite{wpaxos} uses multiple leaders and flexible quorums to reduce wide-area communication costs.
RS-Paxos~\cite{ecpaxos} reduces network and storage overhead by integrating erasure coding with state-machine replication. 
These systems optimize consensus communication patterns, quorum placement, or data-transfer costs. 
In contrast, \sys jointly optimizes proposer rotation and proposal dissemination upon a P2P overlay.

Recent work further explores adaptive consensus under changing network conditions. Crossword~\cite{crossword} dynamically trades off coded-shard assignment and quorum size in response to workload and network conditions, while using lazy follower gossip for failover. Aspen~\cite{aspen} introduces a best-effort sequencing layer based on loosely synchronized clocks and network-delay estimates to accelerate speculative leaderless BFT replication. These approaches reinforce the potential of adapting consensus behavior to observed network conditions, although they target different replication models from rotating-leader consensus.

\noindent\textbf{In-network consensus acceleration.}
Another line of work moves ordering or consensus logic into programmable network devices.
NOPaxos~\cite{nopaxos} replaces coordination on the normal path with an ordered-unreliable-multicast primitive implemented in the network. 
P4xos~\cite{p4xos} implements Paxos logic directly in programmable forwarding devices. 
exposing consensus as a network service. 
Related programmable-network frameworks, such as Emu~\cite{emu}, simplify the prototyping of network services on reconfigurable hardware. 
These approaches achieve substantial acceleration by relying on datacenter-network functionality or programmable devices. 
By contrast, \sys retains host-based consensus and operates over conventional P2P networks without requiring specialized network hardware.

\noindent\textbf{Gossip protocols and topology-aware overlays.}
Gossip-based dissemination has long been studied as a robust communication primitive~\cite{gossiplimits}. 
Its redundancy improves resilience, but it may incur unnecessary bandwidth consumption and propagation delay when directly applied to large proposals. 
Prior work has also investigated whether Internet latency and bandwidth can be approximated using tree-like models~\cite{treeness}. 
These observations motivate the design of \sys.

\section{Conclusion}

We present \sys, a cross-layer design for improving blockchain systems built on P2P networks and rotating-leader consensus.
\sys optimizes the proposer sequence by leveraging the geographic locality of nodes, thereby accelerating proposer rotation.
By leveraging consensus votes as acknowledgments, \sys eliminates redundant proposal transmissions in the normal case while preserving reliable delivery under failures. 
We integrate \sys into Tendermint and libp2p and evaluate our prototype
on Amazon EC2. 
It is promising to further integrate the ideas
of \sys into multi-leader or leaderless protocols.

\bibliographystyle{ACM-Reference-Format}
\bibliography{references}

\clearpage
\appendix

\begingroup
\forcsvlist{\enableappendixfloatcommand}{@dbflt,@dblfloatplacement,@addtodblcol,@addtocurcol,@addtotoporbot,@addtobot,@doclearpage,@combinedblfloats}
\expandafter\let\expandafter\appendixwidealgorithm\csname algorithm*\endcsname
\expandafter\let\expandafter\endappendixwidealgorithm\csname endalgorithm*\endcsname
\newcount\appendixalgorithmcount
\appendixalgorithmcount=0
\renewenvironment{algorithm*}[1][]{%
  \global\advance\appendixalgorithmcount by 1
  \ifnum\appendixalgorithmcount=1
    \appendixwidealgorithm[!b]%
  \else
    \appendixwidealgorithm[!tp]%
  \fi
}{\endappendixwidealgorithm}
\section{Pseudocode of \sys Tendermint}
\label{sec:pesudo}

We present the pseudocode of \sys, adapted from Tendermint~\cite{tendermint}, in Algorithms~\ref{alg:tendermintevent} and~\ref{alg:tendermintfunc}, which describe normal operations and timeout events, respectively.
We highlight the modifications to Tendermint in \algemph{grey}.

\begin{algorithm*}[!ht]
  \caption{\sys (Tendermint) code for node $p$: normal operations.}\label{alg:tendermintevent}
	\scriptsize
  \begin{algorithmic}[1]
	\Statex
	\textbf{Init}:
$h_p\leftarrow 0$, $round_p\leftarrow 0$, $step_p\in\{$\algemph{$tree$,}$propose,prevote,precommit\}$, $decision_p[]\leftarrow nil$, $lockedValue_p\leftarrow nil$, $lockedRound_p\leftarrow -1$, $validValue_p\leftarrow nil$, $validRound_p\leftarrow -1$
  \newline
  \State \textbf{upon} start \textbf{do} $startRound(0)$
  \newline
  \State \textbf{upon} \msg{\proposal}{h_p,round_p,v,-1} from \texttt{proposer}$(h_p,round_p)$ \textbf{while} $step_p=propose$\algemph{$\vee step_p=tree$}  \textbf{do}
  \State \myindent \textbf{if} $valid(v)\wedge(lockedRound_p=-1\vee lockedValue_p=v)$ \textbf{then}
  \State \myindent\myindent \textbf{broadcast} \msg{\prevote}{h_p,round_p,hash(v)}
  \State \myindent \textbf{else}
  \State \myindent\myindent \textbf{broadcast} \msg{\prevote}{h_p,round_p,nil}
  \State \myindent $step_p\leftarrow prevote$
  \newline
  \State \textbf{upon} \msg{\proposal}{h_p,round_p,v,vr} from \texttt{proposer}$(h_p,round_p)$ \textbf{AND} $2f+1$ \msg{prevote}{h_p,vr,hash(v)} \textbf{while} ($step_p=propose$\algemph{$\vee step_p=tree$})$\wedge(vr\geq 0\wedge vr<round_p)$ \textbf{do}
  \State \myindent \textbf{if} $valid(v)\wedge(lockedRound_p\leq vr\vee lockedValue_p=v)$ \textbf{then}
  \State \myindent\myindent \textbf{broadcast} \msg{\prevote}{h_p,round_p,hash(v)}
  \State \myindent \textbf{else}
  \State \myindent\myindent \textbf{broadcast} \msg{\prevote}{h_p,round_p,nil}
  \State \myindent $step_p\leftarrow prevote$
  \newline
  \State \textbf{upon} $2f+1$ \msg{\prevote}{h,r,*} for the first time \textbf{do}
  \State \myindent \textbf{start} $prevoteTimer(h,r)$  \algemph{($3\Delta+2\delta$ time, or $\Delta_L+2\Delta_S+2\delta$)}
  \newline
  \State \textbf{upon} \msg{\proposal}{h_p,round_p,v,*} from \texttt{proposer}$(h_p,round_p)$ \textbf{AND} $2f+1$ \msg{\prevote}{h_p,round_p,hash(v)} \textbf{while} $valid(v)\wedge step_p\geq prevote$ for the first time \textbf{do}
  \State \myindent \textbf{if} $step_p=prevote\wedge \neg timeoutPrevote(h_p,round_p)$ \textbf{then}
  \State \myindent\myindent $lockedValue_p\leftarrow v$
  \State \myindent\myindent $lockedRound_p\leftarrow round_p$
  \State \myindent\myindent \textbf{broadcast} \msg{\precommit}{h_p,round_p,hash(v)}
  \State \myindent\myindent $step_p\leftarrow precommit$
  \State \myindent $validValue_p\leftarrow v$
  \State \myindent $validRound_p\leftarrow round_p$
  \newline
  \State \textbf{upon} $2f+1$ \msg{\precommit}{h_p,round_p,*} \algemph{\textbf{AND} $timeoutPrevote(h_p,round_p)$} for the first time \textbf{do}
  \State \myindent \textbf{start} $precommitTimer(h_p,round_p)$ ($2\Delta$ time, or $2\Delta_S$)
  \newline
  \State \textbf{upon} \msg{\proposal}{h_p,r,v,*} from \texttt{proposer}$(h_p,r)$ \textbf{AND} $2f+1$ \msg{\precommit}{h_p,r,hash(v)} \textbf{while} $decision_p[h_p]=nil$ \textbf{do}
  \State \myindent \textbf{if} $valid(v)$ \textbf{then}
  \State \myindent\myindent $decision_p[h_p]\leftarrow v$
  \State \myindent\myindent $h_p\leftarrow h_p+1$
  \State \myindent\myindent reset $lockedRound_p,lockedValue_p,validRound_p,validValue_p$ to initial values and empty message log
  \State \myindent\myindent $StartRound(0)$
  \newline
  \State \textbf{upon} $f+1$ \msg{\ensuremath{*}}{h_p,round,*,*} with $round>round_p$ \textbf{do}
  \State \myindent $StartRound(round)$
  \newline
  \State \textbf{Function} $startRound(round):$
  \State \myindent $round_p\leftarrow round$
  \State \myindent \algemph{$step_p\leftarrow tree$}
  \State \myindent \textbf{if} $\texttt{proposer}(h_p,round_p)=p$ \textbf{then}
  \State \myindent\myindent \textbf{if} $validValue_p\neq nil$ \textbf{then}
  \State \myindent\myindent\myindent $proposal\leftarrow validValue_p$
  \State \myindent\myindent \textbf{else}
  \State \myindent\myindent\myindent $proposal\leftarrow getValue()$
  \State \myindent\myindent \algemph{\textbf{tree-broadcast} \msg{\proposal}{h_p,round_p,proposal,validRound_p}}
    \State \myindent \algemph{\textbf{start} $treeTimer(h_p,round_p)$ ($2\delta$ time) if not yet started}
  \newline
  \State \algemph{\textbf{upon} $2f+1$ \msg{\precommit}{h,r,hash(v)} for some $v$ for the first time \textbf{do}}
  \State \myindent \algemph{\textbf{start} $treeTimer(h+1,0)$ ($2\delta$ time)}
\end{algorithmic}
\end{algorithm*}

\begin{algorithm*}[!ht]
  \caption{\sys (Tendermint) code for node $p$: timeout events.}\label{alg:tendermintfunc}
	\scriptsize
  \begin{algorithmic}[1]
  \State \algemph{\textbf{upon} $timeoutTree(h_p,round_p)$ \textbf{while} $step_p=tree$ \textbf{do}}
  \State \myindent \algemph{\textbf{start} $proposeTimer(h_p,round_p)$ ($3\Delta$ time, or $2\Delta_L+\Delta_S$)}
  \State \myindent \algemph{$step_p\leftarrow propose$}
  \newline
  \State \textbf{upon} $timeoutPropose(h_p,round_p)$ \textbf{while} $step_p=propose$ \textbf{do}
  \State \myindent \textbf{broadcast} \msg{\prevote}{h_p,round_p,nil}
  \State \myindent $step_p\leftarrow prevote$
  \newline
  \State \textbf{upon} $timeoutPrevote(h_p,round_p)$ while $step_p=prevote$ \textbf{do}
   \State \myindent \textbf{broadcast} \msg{\precommit}{h_p,round_p,nil}
  \State \myindent $step_p\leftarrow precommit$
  \newline
  \State \textbf{upon} $timeoutPrecommit(h_p,round_p)$ while $step_p=precommit$ \textbf{do}
  \State \myindent $startRound(round_p+1)$
  \newline
  \State \algemph{\textbf{upon} \msg{\proposal}{h,r,v,*} from \texttt{proposer}$(h,r)$ \textbf{AND} \algemph{$timeoutTree(h,r)$} for the first time \textbf{do}}
  \State \myindent \algemph{\textbf{for} $\forall p'\in\peer(p)$ \textbf{do}}
  \State \myindent\myindent \algemph{\textbf{if} not received \msg{\prevote}{h,r,hash(v)} signed by $p'$ \textbf{then}}
  \State \myindent\myindent\myindent \algemph{\textbf{send} \msg{\proposal}{h,r,v,*} to $p'$}
  \newline
  \State \algemph{\textbf{upon} $2f+1$ \msg{\prevote}{h_p,round_p,nil} \textbf{do}}
  \State \myindent \algemph{$StartRound(round_p+1)$}
\end{algorithmic}
\end{algorithm*}

\clearpage
\endgroup
\section{Per-Region Node Placement}
\label{app:placement}

Table~\ref{tab:realistic-placement} gives the exact per-region node counts for the Ethereum-realistic deployment of Section~\ref{subsec:hetero}, derived as described there and renormalized to 100 nodes. Counts are taken from Ethernodes~\cite{ethernodes} (by node count) and mapped to the nearest supported AWS region.

\begin{table}[htbp]
  \centering
  \small
  \caption{Per-region node counts for the Ethereum-realistic deployment (100 nodes).}
  \label{tab:realistic-placement}
  \begin{tabular}{lr}
    \toprule
    AWS region (location) & nodes \\
    \midrule
    \multicolumn{2}{l}{\textit{North America}} \\
    \quad \texttt{us-east-1} (Virginia)        & 20 \\
    \quad \texttt{us-east-2} (Ohio)            & 10 \\
    \quad \texttt{us-west-1} (N.\ California)  & 9  \\
    \quad \textit{subtotal}                    & 39 \\
    \midrule
    \multicolumn{2}{l}{\textit{Europe}} \\
    \quad \texttt{eu-central-1} (Frankfurt)    & 30 \\
    \quad \texttt{eu-west-1} (Ireland)         & 12 \\
    \quad \texttt{eu-west-2} (London)          & 4  \\
    \quad \textit{subtotal}                    & 46 \\
    \midrule
    \multicolumn{2}{l}{\textit{Asia-Pacific}} \\
    \quad \texttt{ap-northeast-1} (Tokyo)      & 7  \\
    \quad \texttt{ap-southeast-1} (Singapore)  & 6  \\
    \quad \texttt{ap-southeast-2} (Sydney)     & 2  \\
    \quad \texttt{ap-south-1} (Mumbai)         & 0  \\
    \quad \textit{subtotal}                    & 15 \\
    \midrule
    \textbf{Total}                             & \textbf{100} \\
    \bottomrule
  \end{tabular}
\end{table}

\end{document}